\documentclass[onecolumn, superscriptaddress, amsmath, amssymb, aps, pra]{revtex4-1}

\usepackage{graphicx}
\usepackage{dcolumn}
\usepackage{bm}
\usepackage{ulem}
\usepackage[dvipdfm,colorlinks,linkcolor=blue, urlcolor=red, anchorcolor=blue, citecolor=red]{hyperref}

\begin{document}

\title{Formation of ultracold deeply-bound molecules via multi-state chainwise coincident pulses technique}
\author{Jiahui Zhang}
\email{y10220159@mail.ecust.edu.cn}
\affiliation{School of Physics, East China University of Science and Technology, Shanghai 200237, China}

\begin{abstract}
In this paper, a theoretical method for the efficient creation and detection of deeply bound molecules in three-state $\Lambda$-type and five-state M-type molecular systems is proposed. The method is based on the three-state coincident
pulses technique and the generalized five-state coincident pulses technique. For the three-state system, the technique can efficiently transfer the populations from the Feshbach state to the deeply-bound state via a train of $N$ pairs of resonant and coincident pump and Stokes pulses, with negligible transient populations of excited states. For the five-state system, it is found that this M-type system can be generalized into a $\Lambda$-type structure with the simplest resonant coupling under the assumption of large one-photon detuning together with a requirement of the relation among the four incident pulses. Thereafter, this generalized model permits us to employ the reduced three-state propagator to design four coincident pulses to achieve the desired population transfer. For the numerical study, $^{87}$Rb$_2$ is considered and, it is shown that the weakly-bound Feshbach molecules can be efficiently transferred to their deeply-bound states without strong laser pulses, and the populations of all intermediate states can be well suppressed.
\end{abstract}

\maketitle

\section{\label{sec:level1}Introduction}
Interest in ultracold molecules has grown rapidly in recent years, \cite{10.1063/1.3357286, Q2012} as they offer exciting prospects in areas such as ultracold chemistry, \cite{10.1063/1.4964096} precision measurements, \cite{Ulmanis2012} quantum computation \cite{Carr2009} and quantum simulation. \cite{Covey_2018} But for many of the envisaged studies and applications, initial preparation of the molecular sample in the rovibronic ground state, i.e., the lowest energy level of the electronic ground state, is a prerequisite, which remains a significant challenge. \cite{D1CS01040A}

In the ongoing experiments, ultracold molecules are first created via binding together pre-cooled atoms. This process exploits the existence of scattering resonances, known as Feshbach resonances. \cite{RevModPhys.82.1225} Subsequently, the weakly-bound Feshbach molecules must be quickly moved to their absolute ground state through the well-known STIRAP technique. \cite{PhysRevLett.98.043201, Ospelkaus2008, Danzl1062, doi:10.1126/science.aau7230, 10.1063/1.5108637, 10.1063/5.0046194, 10.1063/5.0082309, PhysRevA.108.043710} During this process, at least one intermediate excited state should be introduced as a bridge between the Feshbach and the ground state, which should have favorable transition dipole moments and good Frank-Condon (FC) overlaps with the vibrational wavefunctions of the Feshbach and deeply-bound ground states. The underlying physical mechanism of STIRAP relies on the existence of a dark state, which should be followed adiabatically. \cite{10.1063/1.458514, 10.1063/1.4916903}
However, the adiabatic requirements are not always affordable in reality, and the highest efficiencies for STIRAP transfer reported hitherto are around $90\%$. \cite{PhysRevLett.113.255301, Christakis2023} High transfer efficiency minimizes the loss in phase-space density when an ensemble
of Feshbach molecule is transferred to the ground state. \cite{Duda2023}

Although many optimization strategies for STIRAP have been developed, \cite{RevModPhys.89.015006, RevModPhys.91.045001, 10.1063/1.4922779} but these come at the expense of strict relations on the pulse shapes, \cite{PhysRevA.80.013417, Zhou2017}, or even require additional couplings between the states that are involved, \cite{doi:10.1063/1.2992152, PhysRevLett.105.123003} which may be impractical for molecules due to different spin characters or weak Franck-Condon overlap between the molecular states that are involved. \cite{Danzl2010, PhysRevA.78.021402, 10.1063/5.0183063, PhysRevA.109.023109} Remarkably, Rangelov and Vitanov have proposed a coincident pulses technique to complete population transfer in three-state systems by a train of $N$ pairs of coincident pulses, \cite{PhysRevA.85.043407} in which the population in the intermediate excited state is suppressed to negligible small value by increasing the pulse pairs. This technique is very attractive because it does not introduce additional couplings, and the pulse shape is not important since the technique uses fields on exact resonance. Recently, the technique has been generalized to tripod system for coherent control of nuclear states. \cite{PhysRevC.96.044619} However, the research on the preparation of ultracold deeply-bound molecules by coincident pulse technique in three-state $\Lambda$-type have not been reported so far. It is also very important to generalize coincident pulse technique into M-type systems, this is due to the fact that the chainwise-STIRAP in M-type molecular system has been proven to be a good alternative in creating ultracold deeply-bound molecules if the typical STIRAP in $\Lambda$-type system does not work due to weak FC factors between the molecular states that are involved. \cite{Danzl2010, PhysRevA.78.021402}

In this paper a theoretical method for the efficient and robust creation of deeply-bound molecules based on the coincident pulses technique is proposed. In the present approach molecules are brought to their deeply-bound ground state through a three-state $\Lambda$-type and a five-state M-type transfer schemes, respectively.
For the three-state transfer scheme, the technique can efficiently transfer the populations from the Feshbach state to the deeply-bound state via a train of $N$ pairs of resonant and coincident pump and Stokes pulses, the population in the intermediate excited state is suppressed to negligible small value by increasing numbers of pulse pairs.
For the five-state transfer scheme, we first reduce the dynamics of the M-type molecular systems to that of effective three-state counterparts under the assumption of large one-photon detuning, by further setting a requirement towards the relation among the original incident pulses, i.e., the pulses at both ends are equal valued and simultaneous, and are the root mean square (rms) of the middle two pulses. In this case, the reduced system can be further generalized into a $\Lambda$-type structure with the simplest resonant coupling. Thereafter, this generalized model permits us to to employ the reduced three-state propagator to design four coincident pulses to achieve the desired population transfer. The calculations reveal that the weakly-bound molecules can be efficiently transferred to their deeply-bound states without strong laser pulses, and the populations of all intermediate states are well suppressed.
\begin{figure}[t]
\centering{\includegraphics[width=6cm]{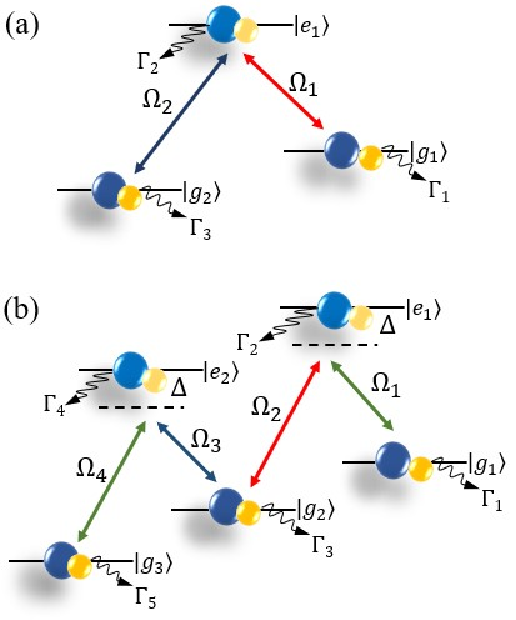}}
\caption{Schematic representations of coherent transfer ultracold molecule via (a) a three-state $\Lambda$-type scheme and (b) a five-state M-type scheme. $\Gamma_i   (i=1, 2, 3, 4, 5)$ represents the phenomenological decay rate due to spontaneous emission and collides of the corresponding state.}
\label{fig1}
\end{figure}
\section{\label{sec:level2}Models, Methods and Results}
\subsection{$\Lambda$-type transfer scheme}
Figure.~\ref{fig1}(a) shows a traditional three-state $\Lambda$-type transfer scheme, in which the initial Feshbach state $|g_1\rangle$ and the intermediate excited state $|e_1\rangle$ are coupled by the pulse with Rabi frequencies $\Omega_1(t)$, the intermediate excited state $|e_1\rangle$ and the final ground rovibrational state $|g_2\rangle$ are coupled by the pulse with
Rabi frequencies $\Omega_2(t)$. Note that the direct coupling between $|g_1\rangle$ and $|g_2\rangle$ is usually not feasible due to their different spin character and a negligible FC factor, and excited state $|e_1\rangle$ should have favourable transition dipole moments and good FC overlaps with both $|g_1\rangle$ and $|g_2\rangle$ vibrational wavefunctions.

The total molecular wave function can be expanded as
\begin{eqnarray}\label{1}
| \psi(t)\rangle=c_{1}(t)| g_1\rangle+c_{2}(t)| e_1\rangle+c_{3}(t)| g_2\rangle,
\end{eqnarray}
the vector $c_{1}(t), c_{2}(t)$ and $c_{3}(t)$ are the probability amplitudes of the corresponding state.
The evolution of the system can be described by the time-dependent Schr\"{o}dinger equation:
\begin{eqnarray}\label{2}
i\hbar\displaystyle\frac{\partial}{\partial t}c(t)=H(t)c(t).
\end{eqnarray}
In the rotating-wave approximation (RWA), the time-dependent Hamiltonian can be written as $(\hbar=1)$:
\begin{equation}\label{3}
H(t)=\frac{1}{2}
\begin{bmatrix}
0&\Omega_{1}&0\\
\Omega_{1}&0&\Omega_{2}\\
0&\Omega_{2}&0\\
\end{bmatrix}.
\end{equation}
In which $\Omega_{1}$ and $\Omega_{2}$ are Rabi frequencies of pump and Stokes laser fields. The null main diagonal elements in the matrix $H(t)$ are due to the assumption of one- and two-photon resonances, which is an essential condition of the standard three-state coincident pulses technique.
Here it is assumed that the two Rabi frequencies are pulse-shaped functions with the same time dependence, but possibly with different amplitudes, i.e.,
\begin{subequations}\label{4}
\begin{align}
\Omega_{1}(t)&=\zeta_1f(t),\\
\Omega_{2}(t)&=\zeta_2f(t).
\end{align}
\end{subequations}
In this case, the Schr\"{o}dinger
equation (\ref{2}) is solved exactly by making a transformation to the so-called bright-dark basis. \cite{PhysRevA.85.043407} The exact propagator is given by
\begin{widetext}
\begin{eqnarray}\label{5}
U(\varphi)=
\left(
\begin{array}{ccc}
1-2\sin^2\varphi\sin^2\frac{A}{4}&-i\sin\varphi\sin\frac{A}{2}&-2\sin2\varphi\sin^2\frac{A}{4}\\
-i\sin\varphi\sin\frac{A}{2}&\cos\frac{A}{2}&-i\cos\varphi\sin\frac{A}{2}\\
-2\sin2\varphi\sin^2\frac{A}{4}&-i\cos\varphi\sin\frac{A}{2}&1-2\cos^2\varphi\sin^2\frac{A}{4}\\
\end{array}
\right),
\end{eqnarray}
\end{widetext}
where $\tan\varphi=\Omega_{1}/\Omega_{2}=\zeta_1/\zeta_2$, the rms pulse area $A$ is defined
as $A=\int^t_{t_i}\sqrt{\Omega^2_{1}(t)+\Omega^2_{2}(t)}dt$.
According to propagator (\ref{5}), one can find the exact analytic solution for any initial condition.

The Hamiltonian $H(t)$, Eq.~(\ref{3}), should drive the initial Feshbach state $|g_1\rangle$ to the target deeply-bound ground state $|g_2\rangle$, the populations at the end of the interaction are
\begin{eqnarray}\label{6}
P_1=|U^N_{11}|^2,\quad
P_2=|U^N_{12}|^2,\quad
P_3=|U^N_{13}|^2.
\end{eqnarray}
Obviously, if $\varphi=\pi/4$ (corresponding to $\zeta_1=\zeta_2$) and the rms pulse area is $A=2\pi$, the population of state $|g_1\rangle$ can be completely transferred to the final state $|g_2\rangle$. However, Eq.~(\ref{6}) implies the intermediate excited state $|e_1\rangle$ will receive a significant transient populations. It is to be expected that the resulting efficiency may be extremely low due to transient excitation of very short lived excited state. \cite{PhysRevLett.125.193201, PhysRevA.96.013406, Zhang_2021} In order to suppress the population of the intermediate excited state, one can use a sequence of $N$ pairs of coincident pulse, each with rms pulse area $A(t)=2\pi$ at the
end of the $k$th step and mixing angles $\varphi_k$, the overall propagator is given by
\begin{eqnarray} \label{7}
U^{(N)}=U(\varphi_N)U(\varphi_{N-1})\cdot\cdot\cdot U(\varphi_{2})U(\varphi_{1}),
\end{eqnarray}
where $\varphi_k$ is given by
\begin{eqnarray} \label{8}
\varphi_k=\frac{(2k-1)\pi}{4N}, k=1,2,...,N.
\end{eqnarray}
As a result, the maximum population of the intermediate excited state $|e_1\rangle$ in the middle of each pulse pair is damped to small values by increasing the number of pulse.

In what follows, let us analyze the formation of $^{87}$Rb$_2$ starting from an initial Feshbach molecular state and taking into account major decay mechanisms to verify the validity of the present protocols. \cite{PhysRevLett.98.043201} Since the excited state of the molecule faces the decays due to spontaneous emission and collisions, and the ground molecular states (for bosonic molecules) experience fast inelastic collisions with the background atoms, the decay of each state shown in Fig.~\ref{fig1}(a) should be considered. Accordingly, the evolution of the system can be governed by the Liouville-von Neumann equation:
\begin{eqnarray}\label{9}
i\hbar\displaystyle\frac{d\rho}{dt}=[\tilde{H}, \rho]+\Gamma\rho,
\end{eqnarray}
in which $\rho$ is the density operator and $\Gamma$ represents phenomenological decay rates.

\begin{figure*}[t]
\centering{\includegraphics[width=12.5cm]{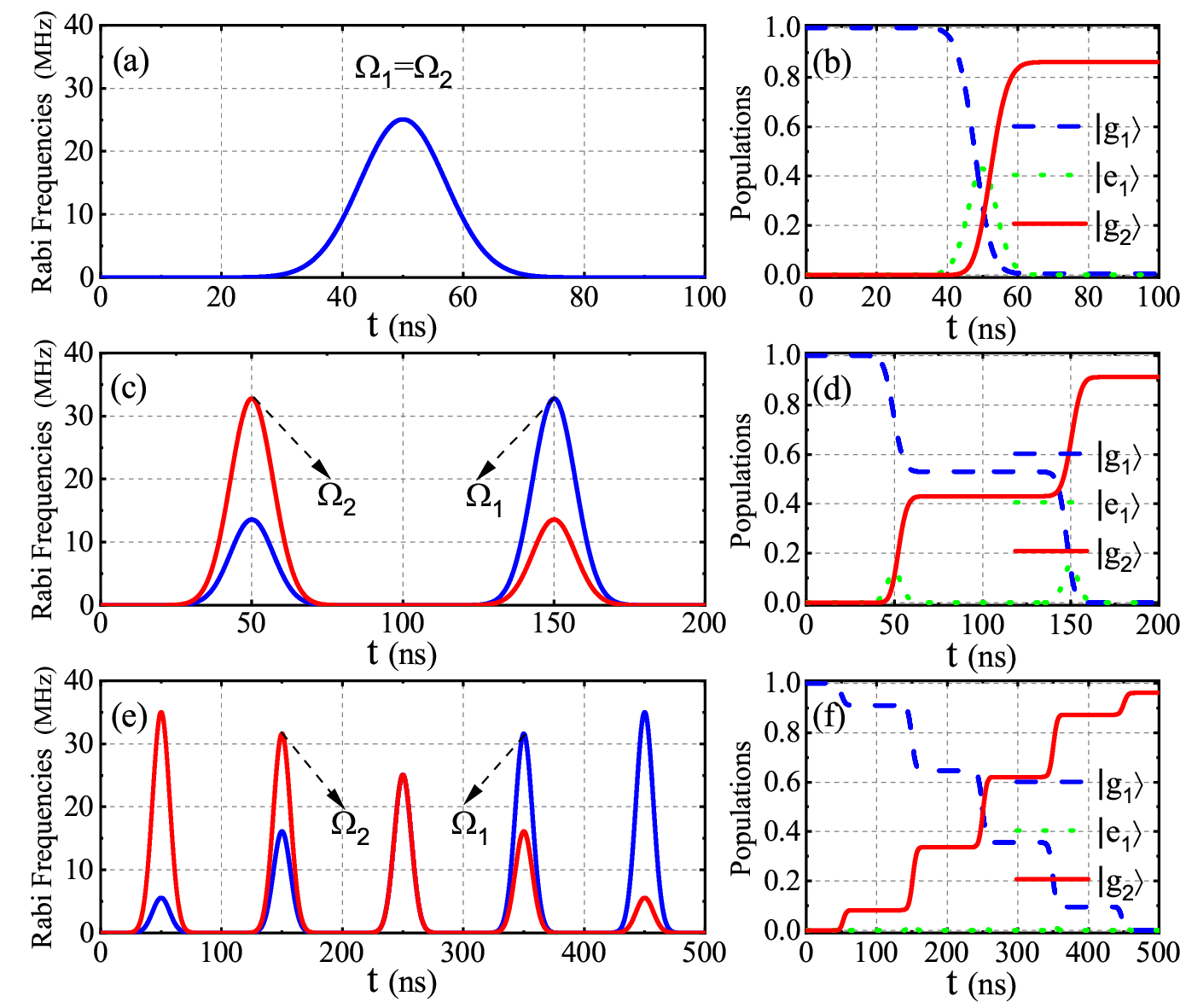}}
\caption{(Color online) Rabi frequencies (left column) and populations (left column) as a function of t for (from top to bottom) $N = 1, 2$, and $5$ pairs of pulses. Here, the two Rabi frequencies of the $k$th are set to be Gaussian with the amplitude envelopes of the forms $\Omega_{1}(t)=\Omega_{0}\sin\varphi_{k}\exp{[-\frac{t-\tau}{t_f}]^2},
\Omega_{2}(t)=\Omega_{0}\cos\varphi_{k}\exp{[-\frac{t-\tau}{t_f}]^2},$ respectively, where $\Omega_{0}= 2\sqrt{\pi}/t_f$ (corresponding to rms pulse area $A=2\pi$) and the mixing angles $\varphi_k (k=1,2,...,5)$ is given by Eq.~(\ref{7}). The parameters
used are $\Gamma_1=0.01{\rm{MHz}}$, $\Gamma_2=30{\rm{MHz}}$, $\Gamma_3=0{\rm{MHz}}$,~\cite{PhysRevA.78.021402} and $t_f=10ns$.}
\label{fig2}
\end{figure*}
The left column of Fig.~\ref{fig2} shows the Rabi frequencies for several pulse trains of different number of pulse pairs, the corresponding the population evolutions are displayed in the right column of Fig.~\ref{fig2}. Clearly, we can see that in all cases the population is transferred from state $|g_1\rangle$ to state $|g_2\rangle$ in the end in a stepwise manner, and the amplitudes of pulses are just around $30\pi{\rm{MHz}}$. As expected, the maximum transient population of the intermediate excited state $|e_1\rangle$ can be damped by increasing the number of pulse pairs. The transfer efficiency of $1, 2$ and $5$ pairs of pulses is $86.13\%$, $91.27\%$ and $97.63\%$ respectively. The above results validate the potential of the present protocol in maintaining the phase-space density of the ultracold gases.

\subsection{M-type transfer scheme}
If the typical STIRAP in $\Lambda$-type system does not work due to the weak FC overlap between the molecular states that are involved, STIRAP in M-type molecular system has been proved to be a good choice for the preparation of ultracold deeply-bound molecules. \cite{Danzl2010, PhysRevA.78.021402, 10.1063/5.0183063, PhysRevA.109.023109}
Figure.~\ref{fig1}(b) shows a five-level M-type transfer scheme, in which both the initial Feshbach state $|g_1\rangle$ and final ground rovibrational state $|g_3\rangle$ are coupled through three intermediate vibrational states. The couplings between states  in the figure is represented by Rabi frequency $\Omega_{j} (j=1, 2, 3, 4)$. Two excited states act as bridges for transfer, one should have a good FC overlap with $|g_3\rangle$, and the other with the initial Feshbach molecular state $|g_1\rangle$. In this case, the two lower states within each sub-$\Lambda$ system are far closer in energy than the initial and final states, thereby greatly boosting the chance of locating an excited state capable of a large FC transition to both lower states.

The total molecular wave function can be expanded as
\begin{eqnarray}\label{10}
| \psi(t)\rangle=c_{1}(t)| g_1\rangle&&+c_{2}(t)| e_1\rangle+c_{3}(t)| g_2\rangle\nonumber \\
&&+c_{4}(t)| e_2\rangle+c_{5}(t)| g_3\rangle.
\end{eqnarray}

The Hamiltonian in the RWA reads$(\hbar=1)$:
\begin{equation}\label{11}
H(t)=
\begin{bmatrix}
0&\Omega_{1}/2&0&0&0\\
\Omega_{1}/2&\Delta&\Omega_{2}/2&0&0\\
0&\Omega_{2}/2&0&\Omega_{3}/2&0\\
0&0&\Omega_{3}/2&\Delta&\Omega_{4}/2\\
0&0&0&\Omega_{4}/2&0\\
\end{bmatrix}.
\end{equation}

When the single-photon detuning $\Delta$ are very large, meaning
$|\Delta_1|\gg\sqrt{\Omega^2_{1}(t)+\Omega^2_{2}(t)}, |\Delta_2|\gg\sqrt{\Omega^2_{3}(t)+\Omega^2_{4}(t)}$), the excited states $| e_{1, 2}\rangle$ can be adiabatically eliminated. Accordingly, Hamiltonian (\ref{11}) can be reduced to the effective form shown below,
\begin{equation}\label{12}
H_e(t)=
\begin{bmatrix}
\Delta_{e_1}&\Omega_{e_1}&0\\
\Omega_{e_1}&\Delta_{e_2}&\Omega_{e_2}\\
0&\Omega_{e_2}&\Delta_{e_3}\\
\end{bmatrix},
\end{equation}
where $\Omega_{e_1}=-\Omega_{1}\Omega_{2}/(4\Delta)$ and $\Omega_{e_2}=-\Omega_{3}\Omega_{4}/(4\Delta)$, and three diagonal elements (dynamic Stark shifts), defined as
$\Delta_{e_1}=-\Omega_{1}^2/(4\Delta),
\Delta_{e_2}=-(\Omega_{2}^2+\Omega_{3}^2)/(4\Delta),
\Delta_{e_3}=-\Omega_{4}^2/(4\Delta)$ respectively. In order to make coincident pulse technique in the strict sense possible, we need to further assume that the three diagonal elements are equal to each other,
i.e., $\Delta_{e_1}=\Delta_{e_2}=\Delta_{e_3}=\Delta'$.
This requires that the four Rabi frequencies should satisfy the following condition:
\begin{eqnarray}\label{13}
\Omega_{1}=\Omega_{4}=\sqrt{\Omega_{2}^2+\Omega_{3}^2}.
\end{eqnarray}
Notably, this kind of pulse sequence is different from previous the straddling-STIRAP~\cite{PhysRevA.56.4929} and the alternating-STIRAP scheme, \cite{PhysRevA.44.7442} which are two possible versions of STIRAP for multilevel systems with odd number of levels.
By setting $c_j =c^{'}_{j}e^{-i\Delta' t}(i=1, 3, 5)$, the Hamiltonian (\ref{12}) can be further equivalent to
\begin{equation}\label{14}
H_e^{'}(t)=
\begin{bmatrix}
0&\Omega^{'}_{e_1}/2&0\\
\Omega^{'}_{e_1}/2&0&\Omega^{'}_{e_2}/2\\
0&\Omega^{'}_{e_2}/2&0\\
\end{bmatrix}.
\end{equation}
where $\Omega^{'}_{e_1}$ and $\Omega^{'}_{e_1}$ represent the two-photon Rabi frequencies, defined as
\begin{subequations}\label{15}
\begin{eqnarray}
\Omega^{'}_{e_1}&=-\frac{\Omega_{2}\sqrt{\Omega_{2}^2+\Omega_{3}^2}}{2\Delta},\\ \Omega^{'}_{e_2}&=-\frac{\Omega_{3}\sqrt{\Omega_{2}^2+\Omega_{3}^2}}{2\Delta},
\end{eqnarray}
\end{subequations}
respectively.
Therefore, we obtain an effective $\Lambda$-type structure with the simplest resonant coupling, as illustrated in Fig.~\ref{fig1}(d).
In principle, the effective Hamiltonian $H_e^{'}(t)$, Eq.~(\ref{14}), can drive the initial Feshbach state $|g_1\rangle$ to the target deeply-bound ground state $|g_3\rangle$ via using the reduced three-state propagator. In order to implement the coincident pulse technique, we impose the condition that the two-photon Rabi frequencies $\Omega^{'}_{e_1}(t)$ and $\Omega^{'}_{e_2}(t)$ are pulse-shaped functions that share the same time dependence, but possibly with different magnitudes. This leads to
\begin{subequations}\label{16}
\begin{eqnarray}
\Omega_{2}(t)&=af(t),\\
\Omega_{3}(t)&=bf(t).
\end{eqnarray}
\end{subequations}
Therefore, we can obtain the reduced three-state propagator, whose form is consistent with the Eq.~(\ref{5}). In this case the angle is redefined as $\tan\varphi=\Omega^{'}_{e_1}/\Omega^{'}_{e_2}=\Omega_{2}/\Omega_{3}=a/b$, and the rms pulse area $A$ is redefined
as $A=\int^t_{t_i}\sqrt{\Omega'^2_{e_1}(t)+\Omega'^2_{e_2}(t)}dt$.

\begin{figure*}[t]
\centering{\includegraphics[width=12.5cm]{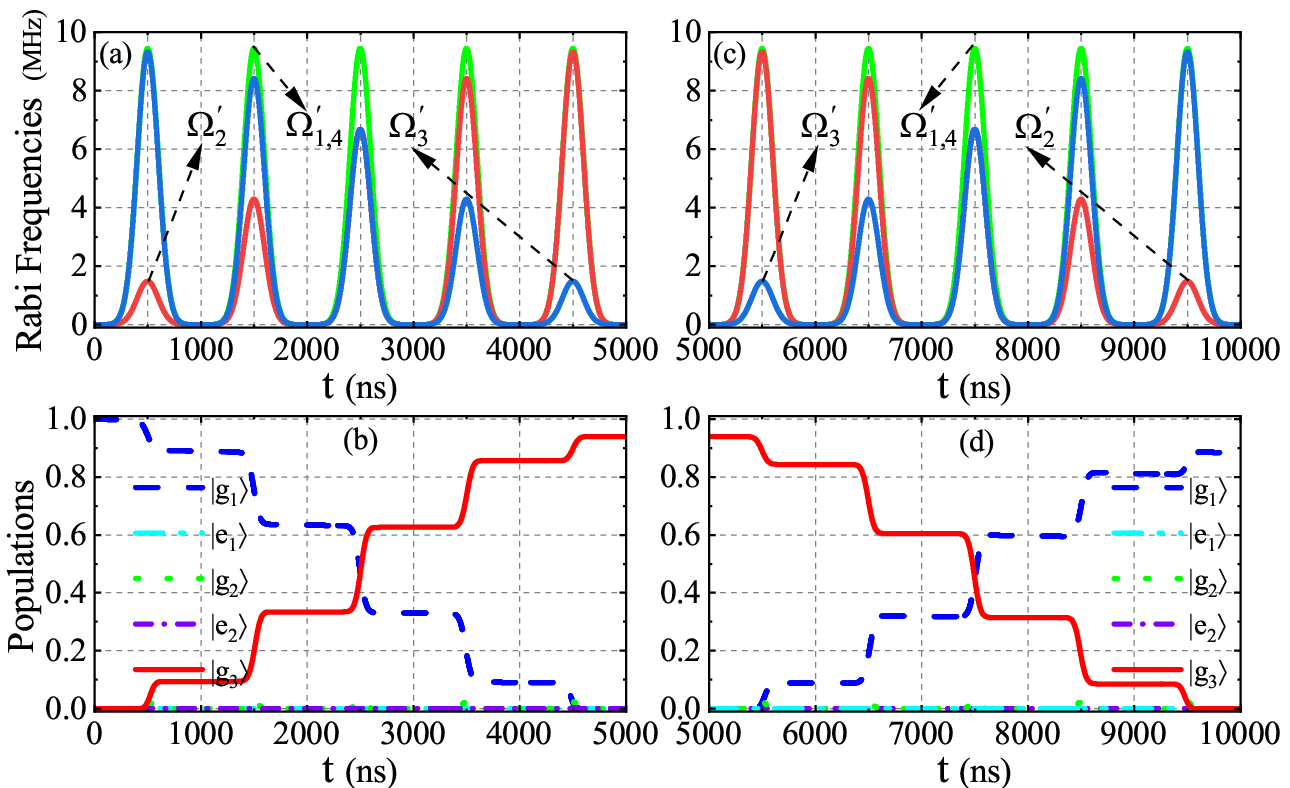}}
\caption{(Color online) Rabi frequencies (left column) and populations (left column) as a function of t for (from top to bottom) $N=2$ pairs of pulses. For convenience, the two-photon Rabi frequencies of the $k$th are set to be Gaussian with the amplitude envelopes of the forms $\Omega^{'}_{e_1}(t)=\Omega_{0}\sin\varphi_{k}\exp{[-\frac{t-\tau}{t_f}]^2},
\Omega^{'}_{e_2}(t)=\Omega_{0}\cos\varphi_{k}\exp{[-\frac{t-\tau}{t_f}]^2},$ respectively, where $\Omega_{0}= 2\sqrt{\pi}/t_f$ (corresponding to rms pulse area $A=2\pi$) and the mixing angles $\varphi_k (k=1,2,...,5)$ is given by Eq.~(\ref{7}). For the Rabi frequencies $\Omega^{'}_{1, 2, 3, 4}$, they are chosen accordingly with respect to Eqs.~(\ref{18}) and~(\ref{19}).}
\label{fig3}
\end{figure*}

According to the above analysis, the reduced system is compatible with three-state coincident pulses technique. However, the two-photon Rabi frequencies for transitions between $|g_1\rangle$ and $|g_2\rangle$ and between $|g_2\rangle$ and $|g_3\rangle$ may be infeasible. \cite{Danzl2010, PhysRevA.78.021402, 10.1063/5.0183063, PhysRevA.109.023109} Therefore, it is necessary to go back to the five-state
structure and design the physically feasible driving fields. Like Eqs.~(\ref{15}), we can impose
\begin{subequations}\label{17}
\begin{eqnarray}
\Omega^{'}_{e_1}&=-\frac{\Omega^{'}_{2}\sqrt{\Omega^{'2}_{2}+\Omega^{'2}_{3}}}{2\Delta^{'}}, \\ \Omega^{'}_{e_2}&=-\frac{\Omega^{'}_{3}\sqrt{\Omega^{'2}_{2}+\Omega^{'2}_{3}}}{2\Delta^{'}}.
\end{eqnarray}
\end{subequations}
By inversely deriving, the explicit expressions of $\Omega^{'}_{2}$ and $\Omega^{'}_{3}$ are obtained as follows:
\begin{subequations}\label{18}
\begin{eqnarray}
\Omega^{'}_{2}&=\Omega^{'}_{e_1}\left(\frac{4\Delta^{'2}}{\Omega^{'2}_{e_1}+\Omega^{'2}_{e_2}}\right)^{1/4},\\
\Omega^{'}_{3}&=\Omega^{'}_{e_2}\left(\frac{4\Delta^{'2}}{\Omega^{'2}_{e_1}+\Omega^{'2}_{e_2}}\right)^{1/4}.
\end{eqnarray}
\end{subequations}
According to Eq.~(\ref{13}), one can impose
$\Omega^{'}_{1, 4}=\sqrt{\Omega^{'2}_{2}+\Omega^{'2}_{3}}$,
and calculate inversely the modified fields as
\begin{eqnarray}\label{19}
\Omega^{'}_{1, 4}=\left[4\Delta^{'2}(\Omega^{'2}_{e_1}+\Omega^{'2}_{e_2}\right]^{1/4}.
\end{eqnarray}
In order to make sure the reduced three-state $\Lambda$-type system can be transformed back to the five-state M-type system, the system must satisfy the adiabatic elimination condition or equivalently $\Delta^{'}\gg\Omega_{j} (j=1, 2, 3, 4)$. Here it is reasonable to assume $\Delta^{'}=\Delta$ since the detuning $\Delta$ is on the order of ${\rm{GHz}}$ while the four Rabi frequencies are on the order of ${\rm{MHz}}$ (see parameters in Figure.~\ref{fig3}).
Therefore, the five-state molecular Hamiltonian can be rewritten as
\begin{equation}\label{20}
H^{'}(t)=
\begin{bmatrix}
0&\Omega^{'}_{1}/2&0&0&0\\
\Omega^{'}_{1}/2&\Delta&\Omega^{'}_{2}/2&0&0\\
0&\Omega^{'}_{2}/2&0&\Omega^{'}_{3}/2&0\\
0&0&\Omega^{'}_{3}/2&\Delta&\Omega^{'}_{4}/2\\
0&0&0&\Omega^{'}_{4}/2&0\\
\end{bmatrix}.
\end{equation}
The above equation implies that we obtain a five-state coincident pulses
technique. We can continue to carry out numerical calculations together with Eq.~(\ref{9}) to verify the feasibility of this protocol. The parameters
used are $\Gamma_1=\Gamma_3=0.01{\rm{MHz}}$, $\Gamma_2=\Gamma_2=30{\rm{MHz}}$, $\Gamma_5=0$,~\cite{PhysRevA.78.021402} $t_f$ is chosen as $100ns$ and $\Delta$ is chosen as $4\pi{\rm{GHz}}$.

As a example, Figure.~\ref{fig3}(a) shows the corresponding time sequence of the four Rabi frequencies for $5$ pairs of pulses, these four coincident pulses are responsible for the distribution of populations from state $| g_1\rangle$ to state $| g_3\rangle$, as illustrated in Fig.~\ref{fig3}(b). The results clearly show the feasibility of the present protocol that high transfer efficiency can be obtained with the peak amplitudes of Rabi frequencies are just around $9.5{\rm{MHz}}$. During the transfer process, the AE protocol ensures the decoupling of the excited states $|e_{1, 2} \rangle$ from the dynamics, which are only used to induce couplings but never significantly populated. Thus, the decay rates of them nearly do not influence the transfer efficiency.

Finally, let us briefly discuss the detection of ultracold deeply-bound molecules. Experimentally, the molecules should be transferred back to the Feshbach state and detected by standard absorption imaging technique. High detection
efficiency is critical for the preparation and characterization of novel many-body phases of dipolar molecules, such as crystalline bulk phases, \cite{PhysRevLett.98.060404} exotic density order, \cite{Baranov2012} or spin order in optical lattices. \cite{PhysRevLett.107.115301}
Generally, the back-transfer is realized by reversing the time sequence of the incident pulses for the creation process.
Figures.~\ref{fig3}(c) and~\ref{fig3}(d) directly gives corresponding examples of the detections following the creation results in Figures.~\ref{fig3}(a) and~\ref{fig3}(b). It is clear that the one-way transfer would nominally convert Feshbach molecules to the rovibrational ground state, while the round-trip transfer would nominally bring the population back.
In these figures, one can clearly see the round-trip transfer of population with perfect efficiencies, which imply high efficiencies of the detection.

\section{\label{sec:level4}SUMMARY}
To conclude, a new theoretical method for the efficient and robust creation and detection of ultracold deeply-bound molecules in a three-state $\Lambda$-type system and in a five-state M-type system have been demonstrated. The method is based on the three-state coincident pulses technique and the generalized five-state coincident pulses technique. Numerical analysis show indeed that both protocols both protocols allow for efficient formation of molecule without strong laser pulses, and the populations of all intermediate states can be well suppressed by increasing the number of pulse pairs. The key of the generalized five-state coincident pulses technique is to first reduce the dynamics into its effective three-state counterpart via adiabatic elimination together with the requirements of the relation among the incident pulses.
Finally, I believe that the proposed protocols could present new possibilities for the creation and detection of deeply-bound molecules.

\nocite{*}
\bibliography{references}

\begin{thebibliography}{42}%
\makeatletter
\providecommand \@ifxundefined [1]{%
 \@ifx{#1\undefined}
}%
\providecommand \@ifnum [1]{%
 \ifnum #1\expandafter \@firstoftwo
 \else \expandafter \@secondoftwo
 \fi
}%
\providecommand \@ifx [1]{%
 \ifx #1\expandafter \@firstoftwo
 \else \expandafter \@secondoftwo
 \fi
}%
\providecommand \natexlab [1]{#1}%
\providecommand \enquote  [1]{``#1''}%
\providecommand \bibnamefont  [1]{#1}%
\providecommand \bibfnamefont [1]{#1}%
\providecommand \citenamefont [1]{#1}%
\providecommand \href@noop [0]{\@secondoftwo}%
\providecommand \href [0]{\begingroup \@sanitize@url \@href}%
\providecommand \@href[1]{\@@startlink{#1}\@@href}%
\providecommand \@@href[1]{\endgroup#1\@@endlink}%
\providecommand \@sanitize@url [0]{\catcode `\\12\catcode `\$12\catcode
  `\&12\catcode `\#12\catcode `\^12\catcode `\_12\catcode `\%12\relax}%
\providecommand \@@startlink[1]{}%
\providecommand \@@endlink[0]{}%
\providecommand \url  [0]{\begingroup\@sanitize@url \@url }%
\providecommand \@url [1]{\endgroup\@href {#1}{\urlprefix }}%
\providecommand \urlprefix  [0]{URL }%
\providecommand \Eprint [0]{\href }%
\providecommand \doibase [0]{http://dx.doi.org/}%
\providecommand \selectlanguage [0]{\@gobble}%
\providecommand \bibinfo  [0]{\@secondoftwo}%
\providecommand \bibfield  [0]{\@secondoftwo}%
\providecommand \translation [1]{[#1]}%
\providecommand \BibitemOpen [0]{}%
\providecommand \bibitemStop [0]{}%
\providecommand \bibitemNoStop [0]{.\EOS\space}%
\providecommand \EOS [0]{\spacefactor3000\relax}%
\providecommand \BibitemShut  [1]{\csname bibitem#1\endcsname}%
\let\auto@bib@innerbib\@empty
\bibitem [{\citenamefont {Chandler}(2010)}]{10.1063/1.3357286}%
  \BibitemOpen
  \bibfield  {author} {\bibinfo {author} {\bibfnamefont {D.~W.}\ \bibnamefont
  {Chandler}},\ }\href {\doibase 10.1063/1.3357286} {\bibfield  {journal}
  {\bibinfo  {journal} {J. Chem. Phys.}\ }\textbf {\bibinfo {volume} {132}},\
  \bibinfo {pages} {110901} (\bibinfo {year} {2010})}\BibitemShut {NoStop}%
\bibitem [{\citenamefont {Qu{\'e}m{\'e}ner}\ and\ \citenamefont
  {Julienne}(2012)}]{Q2012}%
  \BibitemOpen
  \bibfield  {author} {\bibinfo {author} {\bibfnamefont {G.}~\bibnamefont
  {Qu{\'e}m{\'e}ner}}\ and\ \bibinfo {author} {\bibfnamefont {P.~S.}\
  \bibnamefont {Julienne}},\ }\href {\doibase 10.1021/cr300092g} {\bibfield
  {journal} {\bibinfo  {journal} {Chem. Rev.}\ }\textbf {\bibinfo {volume}
  {112}},\ \bibinfo {pages} {4949} (\bibinfo {year} {2012})}\BibitemShut
  {NoStop}%
\bibitem [{\citenamefont {Balakrishnan}(2016)}]{10.1063/1.4964096}%
  \BibitemOpen
  \bibfield  {author} {\bibinfo {author} {\bibfnamefont {N.}~\bibnamefont
  {Balakrishnan}},\ }\href {\doibase 10.1063/1.4964096} {\bibfield  {journal}
  {\bibinfo  {journal} {J. Chem. Phys.}\ }\textbf {\bibinfo {volume} {145}},\
  \bibinfo {pages} {150901} (\bibinfo {year} {2016})}\BibitemShut {NoStop}%
\bibitem [{\citenamefont {Ulmanis}\ \emph {et~al.}(2012)\citenamefont
  {Ulmanis}, \citenamefont {Deiglmayr}, \citenamefont {Repp}, \citenamefont
  {Wester},\ and\ \citenamefont {Weidem{\"u}ller}}]{Ulmanis2012}%
  \BibitemOpen
  \bibfield  {author} {\bibinfo {author} {\bibfnamefont {J.}~\bibnamefont
  {Ulmanis}}, \bibinfo {author} {\bibfnamefont {J.}~\bibnamefont {Deiglmayr}},
  \bibinfo {author} {\bibfnamefont {M.}~\bibnamefont {Repp}}, \bibinfo {author}
  {\bibfnamefont {R.}~\bibnamefont {Wester}}, \ and\ \bibinfo {author}
  {\bibfnamefont {M.}~\bibnamefont {Weidem{\"u}ller}},\ }\href {\doibase
  10.1021/cr300215h} {\bibfield  {journal} {\bibinfo  {journal} {Chem. Rev.}\
  }\textbf {\bibinfo {volume} {112}},\ \bibinfo {pages} {4890} (\bibinfo {year}
  {2012})}\BibitemShut {NoStop}%
\bibitem [{\citenamefont {Carr}\ \emph {et~al.}(2009)\citenamefont {Carr},
  \citenamefont {DeMille}, \citenamefont {Krems},\ and\ \citenamefont
  {Ye}}]{Carr2009}%
  \BibitemOpen
  \bibfield  {author} {\bibinfo {author} {\bibfnamefont {L.~D.}\ \bibnamefont
  {Carr}}, \bibinfo {author} {\bibfnamefont {D.}~\bibnamefont {DeMille}},
  \bibinfo {author} {\bibfnamefont {R.~V.}\ \bibnamefont {Krems}}, \ and\
  \bibinfo {author} {\bibfnamefont {J.}~\bibnamefont {Ye}},\ }\href {\doibase
  10.1088/1367-2630/11/5/055049} {\bibfield  {journal} {\bibinfo  {journal}
  {New J. Phys.}\ }\textbf {\bibinfo {volume} {11}},\ \bibinfo {pages} {055049}
  (\bibinfo {year} {2009})}\BibitemShut {NoStop}%
\bibitem [{\citenamefont {Covey}\ \emph {et~al.}(2018)\citenamefont {Covey},
  \citenamefont {Marco}, \citenamefont {Óscar L~Acevedo}, \citenamefont
  {Rey},\ and\ \citenamefont {Ye}}]{Covey_2018}%
  \BibitemOpen
  \bibfield  {author} {\bibinfo {author} {\bibfnamefont {J.~P.}\ \bibnamefont
  {Covey}}, \bibinfo {author} {\bibfnamefont {L.~D.}\ \bibnamefont {Marco}},
  \bibinfo {author} {\bibnamefont {Óscar L~Acevedo}}, \bibinfo {author}
  {\bibfnamefont {A.~M.}\ \bibnamefont {Rey}}, \ and\ \bibinfo {author}
  {\bibfnamefont {J.}~\bibnamefont {Ye}},\ }\href {\doibase
  10.1088/1367-2630/aaba65} {\bibfield  {journal} {\bibinfo  {journal} {New J.
  Phys.}\ }\textbf {\bibinfo {volume} {20}},\ \bibinfo {pages} {043031}
  (\bibinfo {year} {2018})}\BibitemShut {NoStop}%
\bibitem [{\citenamefont {Zhao}\ and\ \citenamefont {Pan}(2022)}]{D1CS01040A}%
  \BibitemOpen
  \bibfield  {author} {\bibinfo {author} {\bibfnamefont {B.}~\bibnamefont
  {Zhao}}\ and\ \bibinfo {author} {\bibfnamefont {J.-W.}\ \bibnamefont {Pan}},\
  }\href {\doibase 10.1039/D1CS01040A} {\bibfield  {journal} {\bibinfo
  {journal} {Chem. Soc. Rev.}\ }\textbf {\bibinfo {volume} {51}},\ \bibinfo
  {pages} {1685} (\bibinfo {year} {2022})}\BibitemShut {NoStop}%
\bibitem [{\citenamefont {Chin}\ \emph {et~al.}(2010)\citenamefont {Chin},
  \citenamefont {Grimm}, \citenamefont {Julienne},\ and\ \citenamefont
  {Tiesinga}}]{RevModPhys.82.1225}%
  \BibitemOpen
  \bibfield  {author} {\bibinfo {author} {\bibfnamefont {C.}~\bibnamefont
  {Chin}}, \bibinfo {author} {\bibfnamefont {R.}~\bibnamefont {Grimm}},
  \bibinfo {author} {\bibfnamefont {P.}~\bibnamefont {Julienne}}, \ and\
  \bibinfo {author} {\bibfnamefont {E.}~\bibnamefont {Tiesinga}},\ }\href
  {\doibase 10.1103/RevModPhys.82.1225} {\bibfield  {journal} {\bibinfo
  {journal} {Rev. Mod. Phys.}\ }\textbf {\bibinfo {volume} {82}},\ \bibinfo
  {pages} {1225} (\bibinfo {year} {2010})}\BibitemShut {NoStop}%
\bibitem [{\citenamefont {Winkler}\ \emph {et~al.}(2007)\citenamefont
  {Winkler}, \citenamefont {Lang}, \citenamefont {Thalhammer}, \citenamefont
  {Straten}, \citenamefont {Grimm},\ and\ \citenamefont
  {Denschlag}}]{PhysRevLett.98.043201}%
  \BibitemOpen
  \bibfield  {author} {\bibinfo {author} {\bibfnamefont {K.}~\bibnamefont
  {Winkler}}, \bibinfo {author} {\bibfnamefont {F.}~\bibnamefont {Lang}},
  \bibinfo {author} {\bibfnamefont {G.}~\bibnamefont {Thalhammer}}, \bibinfo
  {author} {\bibfnamefont {P.~v.~d.}\ \bibnamefont {Straten}}, \bibinfo
  {author} {\bibfnamefont {R.}~\bibnamefont {Grimm}}, \ and\ \bibinfo {author}
  {\bibfnamefont {J.~H.}\ \bibnamefont {Denschlag}},\ }\href {\doibase
  10.1103/PhysRevLett.98.043201} {\bibfield  {journal} {\bibinfo  {journal}
  {Phys. Rev. Lett.}\ }\textbf {\bibinfo {volume} {98}},\ \bibinfo {pages}
  {043201} (\bibinfo {year} {2007})}\BibitemShut {NoStop}%
\bibitem [{\citenamefont {Ospelkaus}\ \emph {et~al.}(2008)\citenamefont
  {Ospelkaus}, \citenamefont {Pe'er}, \citenamefont {Ni}, \citenamefont
  {Zirbel}, \citenamefont {Neyenhuis}, \citenamefont {Kotochigova},
  \citenamefont {Julienne}, \citenamefont {Ye},\ and\ \citenamefont
  {Jin}}]{Ospelkaus2008}%
  \BibitemOpen
  \bibfield  {author} {\bibinfo {author} {\bibfnamefont {S.}~\bibnamefont
  {Ospelkaus}}, \bibinfo {author} {\bibfnamefont {A.}~\bibnamefont {Pe'er}},
  \bibinfo {author} {\bibfnamefont {K.-K.}\ \bibnamefont {Ni}}, \bibinfo
  {author} {\bibfnamefont {J.~J.}\ \bibnamefont {Zirbel}}, \bibinfo {author}
  {\bibfnamefont {B.}~\bibnamefont {Neyenhuis}}, \bibinfo {author}
  {\bibfnamefont {S.}~\bibnamefont {Kotochigova}}, \bibinfo {author}
  {\bibfnamefont {P.~S.}\ \bibnamefont {Julienne}}, \bibinfo {author}
  {\bibfnamefont {J.}~\bibnamefont {Ye}}, \ and\ \bibinfo {author}
  {\bibfnamefont {D.~S.}\ \bibnamefont {Jin}},\ }\href {\doibase
  10.1038/nphys997} {\bibfield  {journal} {\bibinfo  {journal} {Nat. Phys.}\
  }\textbf {\bibinfo {volume} {4}},\ \bibinfo {pages} {622} (\bibinfo {year}
  {2008})}\BibitemShut {NoStop}%
\bibitem [{\citenamefont {Danzl}\ \emph {et~al.}(2008)\citenamefont {Danzl},
  \citenamefont {Haller}, \citenamefont {Gustavsson}, \citenamefont {Mark},
  \citenamefont {Hart}, \citenamefont {Bouloufa}, \citenamefont {Dulieu},
  \citenamefont {Ritsch},\ and\ \citenamefont {N{\"a}gerl}}]{Danzl1062}%
  \BibitemOpen
  \bibfield  {author} {\bibinfo {author} {\bibfnamefont {J.~G.}\ \bibnamefont
  {Danzl}}, \bibinfo {author} {\bibfnamefont {E.}~\bibnamefont {Haller}},
  \bibinfo {author} {\bibfnamefont {M.}~\bibnamefont {Gustavsson}}, \bibinfo
  {author} {\bibfnamefont {M.~J.}\ \bibnamefont {Mark}}, \bibinfo {author}
  {\bibfnamefont {R.}~\bibnamefont {Hart}}, \bibinfo {author} {\bibfnamefont
  {N.}~\bibnamefont {Bouloufa}}, \bibinfo {author} {\bibfnamefont
  {O.}~\bibnamefont {Dulieu}}, \bibinfo {author} {\bibfnamefont
  {H.}~\bibnamefont {Ritsch}}, \ and\ \bibinfo {author} {\bibfnamefont {H.-C.}\
  \bibnamefont {N{\"a}gerl}},\ }\href {\doibase 10.1126/science.1159909}
  {\bibfield  {journal} {\bibinfo  {journal} {Science}\ }\textbf {\bibinfo
  {volume} {321}},\ \bibinfo {pages} {1062} (\bibinfo {year}
  {2008})}\BibitemShut {NoStop}%
\bibitem [{\citenamefont {Marco}\ \emph {et~al.}(2019)\citenamefont {Marco},
  \citenamefont {Valtolina}, \citenamefont {Matsuda}, \citenamefont {Tobias},
  \citenamefont {Covey},\ and\ \citenamefont
  {Ye}}]{doi:10.1126/science.aau7230}%
  \BibitemOpen
  \bibfield  {author} {\bibinfo {author} {\bibfnamefont {L.~D.}\ \bibnamefont
  {Marco}}, \bibinfo {author} {\bibfnamefont {G.}~\bibnamefont {Valtolina}},
  \bibinfo {author} {\bibfnamefont {K.}~\bibnamefont {Matsuda}}, \bibinfo
  {author} {\bibfnamefont {W.~G.}\ \bibnamefont {Tobias}}, \bibinfo {author}
  {\bibfnamefont {J.~P.}\ \bibnamefont {Covey}}, \ and\ \bibinfo {author}
  {\bibfnamefont {J.}~\bibnamefont {Ye}},\ }\href {\doibase
  10.1126/science.aau7230} {\bibfield  {journal} {\bibinfo  {journal}
  {Science}\ }\textbf {\bibinfo {volume} {363}},\ \bibinfo {pages} {853}
  (\bibinfo {year} {2019})}\BibitemShut {NoStop}%
\bibitem [{\citenamefont {Liu}\ \emph {et~al.}(2019)\citenamefont {Liu},
  \citenamefont {Gong}, \citenamefont {Ji}, \citenamefont {Wang}, \citenamefont
  {Zhao}, \citenamefont {Xiao},\ and\ \citenamefont {Jia}}]{10.1063/1.5108637}%
  \BibitemOpen
  \bibfield  {author} {\bibinfo {author} {\bibfnamefont {Y.}~\bibnamefont
  {Liu}}, \bibinfo {author} {\bibfnamefont {T.}~\bibnamefont {Gong}}, \bibinfo
  {author} {\bibfnamefont {Z.}~\bibnamefont {Ji}}, \bibinfo {author}
  {\bibfnamefont {G.}~\bibnamefont {Wang}}, \bibinfo {author} {\bibfnamefont
  {Y.}~\bibnamefont {Zhao}}, \bibinfo {author} {\bibfnamefont {L.}~\bibnamefont
  {Xiao}}, \ and\ \bibinfo {author} {\bibfnamefont {S.}~\bibnamefont {Jia}},\
  }\href {\doibase 10.1063/1.5108637} {\bibfield  {journal} {\bibinfo
  {journal} {J. Chem. Phys.}\ }\textbf {\bibinfo {volume} {151}},\ \bibinfo
  {pages} {084303} (\bibinfo {year} {2019})}\BibitemShut {NoStop}%
\bibitem [{\citenamefont {Schwarzer}\ and\ \citenamefont
  {Toennies}(2021)}]{10.1063/5.0046194}%
  \BibitemOpen
  \bibfield  {author} {\bibinfo {author} {\bibfnamefont {M.}~\bibnamefont
  {Schwarzer}}\ and\ \bibinfo {author} {\bibfnamefont {J.~P.}\ \bibnamefont
  {Toennies}},\ }\href {\doibase 10.1063/5.0046194} {\bibfield  {journal}
  {\bibinfo  {journal} {J. Chem. Phys.}\ }\textbf {\bibinfo {volume} {154}},\
  \bibinfo {pages} {154304} (\bibinfo {year} {2021})}\BibitemShut {NoStop}%
\bibitem [{\citenamefont {Krumins}\ \emph {et~al.}(2022)\citenamefont
  {Krumins}, \citenamefont {Kruzins}, \citenamefont {Tamanis}, \citenamefont
  {Ferber}, \citenamefont {Meshkov}, \citenamefont {Pazyuk}, \citenamefont
  {Stolyarov},\ and\ \citenamefont {Pashov}}]{10.1063/5.0082309}%
  \BibitemOpen
  \bibfield  {author} {\bibinfo {author} {\bibfnamefont {V.}~\bibnamefont
  {Krumins}}, \bibinfo {author} {\bibfnamefont {A.}~\bibnamefont {Kruzins}},
  \bibinfo {author} {\bibfnamefont {M.}~\bibnamefont {Tamanis}}, \bibinfo
  {author} {\bibfnamefont {R.}~\bibnamefont {Ferber}}, \bibinfo {author}
  {\bibfnamefont {V.~V.}\ \bibnamefont {Meshkov}}, \bibinfo {author}
  {\bibfnamefont {E.~A.}\ \bibnamefont {Pazyuk}}, \bibinfo {author}
  {\bibfnamefont {A.~V.}\ \bibnamefont {Stolyarov}}, \ and\ \bibinfo {author}
  {\bibfnamefont {A.}~\bibnamefont {Pashov}},\ }\href {\doibase
  10.1063/5.0082309} {\bibfield  {journal} {\bibinfo  {journal} {J. Chem.
  Phys.}\ }\textbf {\bibinfo {volume} {156}},\ \bibinfo {pages} {114305}
  (\bibinfo {year} {2022})}\BibitemShut {NoStop}%
\bibitem [{\citenamefont {Chathanathil}\ \emph {et~al.}(2023)\citenamefont
  {Chathanathil}, \citenamefont {Ramaswamy}, \citenamefont {Malinovsky},
  \citenamefont {Budker},\ and\ \citenamefont
  {Malinovskaya}}]{PhysRevA.108.043710}%
  \BibitemOpen
  \bibfield  {author} {\bibinfo {author} {\bibfnamefont {J.}~\bibnamefont
  {Chathanathil}}, \bibinfo {author} {\bibfnamefont {A.}~\bibnamefont
  {Ramaswamy}}, \bibinfo {author} {\bibfnamefont {V.~S.}\ \bibnamefont
  {Malinovsky}}, \bibinfo {author} {\bibfnamefont {D.}~\bibnamefont {Budker}},
  \ and\ \bibinfo {author} {\bibfnamefont {S.~A.}\ \bibnamefont
  {Malinovskaya}},\ }\href {\doibase 10.1103/PhysRevA.108.043710} {\bibfield
  {journal} {\bibinfo  {journal} {Phys. Rev. A}\ }\textbf {\bibinfo {volume}
  {108}},\ \bibinfo {pages} {043710} (\bibinfo {year} {2023})}\BibitemShut
  {NoStop}%
\bibitem [{\citenamefont {Gaubatz}\ \emph {et~al.}(1990)\citenamefont
  {Gaubatz}, \citenamefont {Rudecki}, \citenamefont {Schiemann},\ and\
  \citenamefont {Bergmann}}]{10.1063/1.458514}%
  \BibitemOpen
  \bibfield  {author} {\bibinfo {author} {\bibfnamefont {U.}~\bibnamefont
  {Gaubatz}}, \bibinfo {author} {\bibfnamefont {P.}~\bibnamefont {Rudecki}},
  \bibinfo {author} {\bibfnamefont {S.}~\bibnamefont {Schiemann}}, \ and\
  \bibinfo {author} {\bibfnamefont {K.}~\bibnamefont {Bergmann}},\ }\href
  {\doibase 10.1063/1.458514} {\bibfield  {journal} {\bibinfo  {journal} {J.
  Chem. Phys.}\ }\textbf {\bibinfo {volume} {92}},\ \bibinfo {pages} {5363}
  (\bibinfo {year} {1990})}\BibitemShut {NoStop}%
\bibitem [{\citenamefont {Bergmann}\ \emph {et~al.}(2015)\citenamefont
  {Bergmann}, \citenamefont {Vitanov},\ and\ \citenamefont
  {Shore}}]{10.1063/1.4916903}%
  \BibitemOpen
  \bibfield  {author} {\bibinfo {author} {\bibfnamefont {K.}~\bibnamefont
  {Bergmann}}, \bibinfo {author} {\bibfnamefont {N.~V.}\ \bibnamefont
  {Vitanov}}, \ and\ \bibinfo {author} {\bibfnamefont {B.~W.}\ \bibnamefont
  {Shore}},\ }\href {\doibase 10.1063/1.4916903} {\bibfield  {journal}
  {\bibinfo  {journal} {J. Chem. Phys.}\ }\textbf {\bibinfo {volume} {142}},\
  \bibinfo {pages} {170901} (\bibinfo {year} {2015})}\BibitemShut {NoStop}%
\bibitem [{\citenamefont {Molony}\ \emph {et~al.}(2014)\citenamefont {Molony},
  \citenamefont {Gregory}, \citenamefont {Ji}, \citenamefont {Lu},
  \citenamefont {K{\"o}ppinger}, \citenamefont {Le~Sueur}, \citenamefont
  {Blackley}, \citenamefont {Hutson},\ and\ \citenamefont
  {Cornish}}]{PhysRevLett.113.255301}%
  \BibitemOpen
  \bibfield  {author} {\bibinfo {author} {\bibfnamefont {P.~K.}\ \bibnamefont
  {Molony}}, \bibinfo {author} {\bibfnamefont {P.~D.}\ \bibnamefont {Gregory}},
  \bibinfo {author} {\bibfnamefont {Z.}~\bibnamefont {Ji}}, \bibinfo {author}
  {\bibfnamefont {B.}~\bibnamefont {Lu}}, \bibinfo {author} {\bibfnamefont
  {M.~P.}\ \bibnamefont {K{\"o}ppinger}}, \bibinfo {author} {\bibfnamefont
  {C.~R.}\ \bibnamefont {Le~Sueur}}, \bibinfo {author} {\bibfnamefont {C.~L.}\
  \bibnamefont {Blackley}}, \bibinfo {author} {\bibfnamefont {J.~M.}\
  \bibnamefont {Hutson}}, \ and\ \bibinfo {author} {\bibfnamefont {S.~L.}\
  \bibnamefont {Cornish}},\ }\href {\doibase 10.1103/PhysRevLett.113.255301}
  {\bibfield  {journal} {\bibinfo  {journal} {Phys. Rev. Lett.}\ }\textbf
  {\bibinfo {volume} {113}},\ \bibinfo {pages} {255301} (\bibinfo {year}
  {2014})}\BibitemShut {NoStop}%
\bibitem [{\citenamefont {Christakis}\ \emph {et~al.}(2023)\citenamefont
  {Christakis}, \citenamefont {Rosenberg}, \citenamefont {Raj}, \citenamefont
  {Chi}, \citenamefont {Morningstar}, \citenamefont {Huse}, \citenamefont
  {Yan},\ and\ \citenamefont {Bakr}}]{Christakis2023}%
  \BibitemOpen
  \bibfield  {author} {\bibinfo {author} {\bibfnamefont {L.}~\bibnamefont
  {Christakis}}, \bibinfo {author} {\bibfnamefont {J.~S.}\ \bibnamefont
  {Rosenberg}}, \bibinfo {author} {\bibfnamefont {R.}~\bibnamefont {Raj}},
  \bibinfo {author} {\bibfnamefont {S.}~\bibnamefont {Chi}}, \bibinfo {author}
  {\bibfnamefont {A.}~\bibnamefont {Morningstar}}, \bibinfo {author}
  {\bibfnamefont {D.~A.}\ \bibnamefont {Huse}}, \bibinfo {author}
  {\bibfnamefont {Z.~Z.}\ \bibnamefont {Yan}}, \ and\ \bibinfo {author}
  {\bibfnamefont {W.~S.}\ \bibnamefont {Bakr}},\ }\href {\doibase
  10.1038/s41586-022-05558-4} {\bibfield  {journal} {\bibinfo  {journal}
  {Nature}\ }\textbf {\bibinfo {volume} {614}},\ \bibinfo {pages} {64}
  (\bibinfo {year} {2023})}\BibitemShut {NoStop}%
\bibitem [{\citenamefont {Duda}\ \emph {et~al.}(2023)\citenamefont {Duda},
  \citenamefont {Chen}, \citenamefont {Schindewolf}, \citenamefont {Bause},
  \citenamefont {von Milczewski}, \citenamefont {Schmidt}, \citenamefont
  {Bloch},\ and\ \citenamefont {Luo}}]{Duda2023}%
  \BibitemOpen
  \bibfield  {author} {\bibinfo {author} {\bibfnamefont {M.}~\bibnamefont
  {Duda}}, \bibinfo {author} {\bibfnamefont {X.-Y.}\ \bibnamefont {Chen}},
  \bibinfo {author} {\bibfnamefont {A.}~\bibnamefont {Schindewolf}}, \bibinfo
  {author} {\bibfnamefont {R.}~\bibnamefont {Bause}}, \bibinfo {author}
  {\bibfnamefont {J.}~\bibnamefont {von Milczewski}}, \bibinfo {author}
  {\bibfnamefont {R.}~\bibnamefont {Schmidt}}, \bibinfo {author} {\bibfnamefont
  {I.}~\bibnamefont {Bloch}}, \ and\ \bibinfo {author} {\bibfnamefont {X.-Y.}\
  \bibnamefont {Luo}},\ }\href {\doibase 10.1038/s41567-023-01948-1} {\bibfield
   {journal} {\bibinfo  {journal} {Nat. Phys.}\ }\textbf {\bibinfo {volume}
  {19}},\ \bibinfo {pages} {720} (\bibinfo {year} {2023})}\BibitemShut
  {NoStop}%
\bibitem [{\citenamefont {Vitanov}\ \emph {et~al.}(2017)\citenamefont
  {Vitanov}, \citenamefont {Rangelov}, \citenamefont {Shore},\ and\
  \citenamefont {Bergmann}}]{RevModPhys.89.015006}%
  \BibitemOpen
  \bibfield  {author} {\bibinfo {author} {\bibfnamefont {N.~V.}\ \bibnamefont
  {Vitanov}}, \bibinfo {author} {\bibfnamefont {A.~A.}\ \bibnamefont
  {Rangelov}}, \bibinfo {author} {\bibfnamefont {B.~W.}\ \bibnamefont {Shore}},
  \ and\ \bibinfo {author} {\bibfnamefont {K.}~\bibnamefont {Bergmann}},\
  }\href {\doibase 10.1103/RevModPhys.89.015006} {\bibfield  {journal}
  {\bibinfo  {journal} {Rev. Mod. Phys.}\ }\textbf {\bibinfo {volume} {89}},\
  \bibinfo {pages} {015006} (\bibinfo {year} {2017})}\BibitemShut {NoStop}%
\bibitem [{\citenamefont {Gu\'ery-Odelin}\ \emph {et~al.}(2019)\citenamefont
  {Gu\'ery-Odelin}, \citenamefont {Ruschhaupt}, \citenamefont {Kiely},
  \citenamefont {Torrontegui}, \citenamefont {Mart\'{\i}nez-Garaot},\ and\
  \citenamefont {Muga}}]{RevModPhys.91.045001}%
  \BibitemOpen
  \bibfield  {author} {\bibinfo {author} {\bibfnamefont {D.}~\bibnamefont
  {Gu\'ery-Odelin}}, \bibinfo {author} {\bibfnamefont {A.}~\bibnamefont
  {Ruschhaupt}}, \bibinfo {author} {\bibfnamefont {A.}~\bibnamefont {Kiely}},
  \bibinfo {author} {\bibfnamefont {E.}~\bibnamefont {Torrontegui}}, \bibinfo
  {author} {\bibfnamefont {S.}~\bibnamefont {Mart\'{\i}nez-Garaot}}, \ and\
  \bibinfo {author} {\bibfnamefont {J.~G.}\ \bibnamefont {Muga}},\ }\href
  {\doibase 10.1103/RevModPhys.91.045001} {\bibfield  {journal} {\bibinfo
  {journal} {Rev. Mod. Phys.}\ }\textbf {\bibinfo {volume} {91}},\ \bibinfo
  {pages} {045001} (\bibinfo {year} {2019})}\BibitemShut {NoStop}%
\bibitem [{\citenamefont {Masuda}\ and\ \citenamefont
  {Rice}(2015)}]{10.1063/1.4922779}%
  \BibitemOpen
  \bibfield  {author} {\bibinfo {author} {\bibfnamefont {S.}~\bibnamefont
  {Masuda}}\ and\ \bibinfo {author} {\bibfnamefont {S.~A.}\ \bibnamefont
  {Rice}},\ }\href {\doibase 10.1063/1.4922779} {\bibfield  {journal} {\bibinfo
   {journal} {J. Chem. Phys.}\ }\textbf {\bibinfo {volume} {142}},\ \bibinfo
  {pages} {244303} (\bibinfo {year} {2015})}\BibitemShut {NoStop}%
\bibitem [{\citenamefont {Vasilev}\ \emph {et~al.}(2009)\citenamefont
  {Vasilev}, \citenamefont {Kuhn},\ and\ \citenamefont
  {Vitanov}}]{PhysRevA.80.013417}%
  \BibitemOpen
  \bibfield  {author} {\bibinfo {author} {\bibfnamefont {G.~S.}\ \bibnamefont
  {Vasilev}}, \bibinfo {author} {\bibfnamefont {A.}~\bibnamefont {Kuhn}}, \
  and\ \bibinfo {author} {\bibfnamefont {N.~V.}\ \bibnamefont {Vitanov}},\
  }\href {\doibase 10.1103/PhysRevA.80.013417} {\bibfield  {journal} {\bibinfo
  {journal} {Phys. Rev. A}\ }\textbf {\bibinfo {volume} {80}},\ \bibinfo
  {pages} {013417} (\bibinfo {year} {2009})}\BibitemShut {NoStop}%
\bibitem [{\citenamefont {Zhou}\ \emph {et~al.}(2017)\citenamefont {Zhou},
  \citenamefont {Baksic}, \citenamefont {Ribeiro}, \citenamefont {Yale},
  \citenamefont {Heremans}, \citenamefont {Jerger}, \citenamefont {Auer},
  \citenamefont {Burkard}, \citenamefont {Clerk},\ and\ \citenamefont
  {Awschalom}}]{Zhou2017}%
  \BibitemOpen
  \bibfield  {author} {\bibinfo {author} {\bibfnamefont {B.}~\bibnamefont
  {Zhou}}, \bibinfo {author} {\bibfnamefont {A.}~\bibnamefont {Baksic}},
  \bibinfo {author} {\bibfnamefont {H.}~\bibnamefont {Ribeiro}}, \bibinfo
  {author} {\bibfnamefont {C.}~\bibnamefont {Yale}}, \bibinfo {author}
  {\bibfnamefont {F.}~\bibnamefont {Heremans}}, \bibinfo {author}
  {\bibfnamefont {P.}~\bibnamefont {Jerger}}, \bibinfo {author} {\bibfnamefont
  {A.}~\bibnamefont {Auer}}, \bibinfo {author} {\bibfnamefont {G.}~\bibnamefont
  {Burkard}}, \bibinfo {author} {\bibfnamefont {A.}~\bibnamefont {Clerk}}, \
  and\ \bibinfo {author} {\bibfnamefont {D.}~\bibnamefont {Awschalom}},\ }\href
  {\doibase 10.1038/nphys3967} {\bibfield  {journal} {\bibinfo  {journal} {Nat.
  Phys.}\ }\textbf {\bibinfo {volume} {13}},\ \bibinfo {pages} {330} (\bibinfo
  {year} {2017})}\BibitemShut {NoStop}%
\bibitem [{\citenamefont {Demirplak}\ and\ \citenamefont
  {Rice}(2008)}]{doi:10.1063/1.2992152}%
  \BibitemOpen
  \bibfield  {author} {\bibinfo {author} {\bibfnamefont {M.}~\bibnamefont
  {Demirplak}}\ and\ \bibinfo {author} {\bibfnamefont {S.~A.}\ \bibnamefont
  {Rice}},\ }\href {\doibase 10.1063/1.2992152} {\bibfield  {journal} {\bibinfo
   {journal} {J. Chem. Phys.}\ }\textbf {\bibinfo {volume} {129}},\ \bibinfo
  {pages} {154111} (\bibinfo {year} {2008})}\BibitemShut {NoStop}%
\bibitem [{\citenamefont {Chen}\ \emph {et~al.}(2010)\citenamefont {Chen},
  \citenamefont {Lizuain}, \citenamefont {Ruschhaupt}, \citenamefont
  {Gu\'ery-Odelin},\ and\ \citenamefont {Muga}}]{PhysRevLett.105.123003}%
  \BibitemOpen
  \bibfield  {author} {\bibinfo {author} {\bibfnamefont {X.}~\bibnamefont
  {Chen}}, \bibinfo {author} {\bibfnamefont {I.}~\bibnamefont {Lizuain}},
  \bibinfo {author} {\bibfnamefont {A.}~\bibnamefont {Ruschhaupt}}, \bibinfo
  {author} {\bibfnamefont {D.}~\bibnamefont {Gu\'ery-Odelin}}, \ and\ \bibinfo
  {author} {\bibfnamefont {J.~G.}\ \bibnamefont {Muga}},\ }\href {\doibase
  10.1103/PhysRevLett.105.123003} {\bibfield  {journal} {\bibinfo  {journal}
  {Phys. Rev. Lett.}\ }\textbf {\bibinfo {volume} {105}},\ \bibinfo {pages}
  {123003} (\bibinfo {year} {2010})}\BibitemShut {NoStop}%
\bibitem [{\citenamefont {Danzl}\ \emph {et~al.}(2010)\citenamefont {Danzl},
  \citenamefont {Mark}, \citenamefont {Haller}, \citenamefont {Gustavsson},
  \citenamefont {Hart}, \citenamefont {Aldegunde}, \citenamefont {Hutson},\
  and\ \citenamefont {N{\"a}gerl}}]{Danzl2010}%
  \BibitemOpen
  \bibfield  {author} {\bibinfo {author} {\bibfnamefont {J.~G.}\ \bibnamefont
  {Danzl}}, \bibinfo {author} {\bibfnamefont {M.~J.}\ \bibnamefont {Mark}},
  \bibinfo {author} {\bibfnamefont {E.}~\bibnamefont {Haller}}, \bibinfo
  {author} {\bibfnamefont {M.}~\bibnamefont {Gustavsson}}, \bibinfo {author}
  {\bibfnamefont {R.}~\bibnamefont {Hart}}, \bibinfo {author} {\bibfnamefont
  {J.}~\bibnamefont {Aldegunde}}, \bibinfo {author} {\bibfnamefont {J.~M.}\
  \bibnamefont {Hutson}}, \ and\ \bibinfo {author} {\bibfnamefont {H.-C.}\
  \bibnamefont {N{\"a}gerl}},\ }\href {\doibase 10.1038/nphys1533} {\bibfield
  {journal} {\bibinfo  {journal} {Nat. Phys.}\ }\textbf {\bibinfo {volume}
  {6}},\ \bibinfo {pages} {265} (\bibinfo {year} {2010})}\BibitemShut {NoStop}%
\bibitem [{\citenamefont {Kuznetsova}\ \emph {et~al.}(2008)\citenamefont
  {Kuznetsova}, \citenamefont {Pellegrini}, \citenamefont {C\^ot\'e},
  \citenamefont {Lukin},\ and\ \citenamefont {Yelin}}]{PhysRevA.78.021402}%
  \BibitemOpen
  \bibfield  {author} {\bibinfo {author} {\bibfnamefont {E.}~\bibnamefont
  {Kuznetsova}}, \bibinfo {author} {\bibfnamefont {P.}~\bibnamefont
  {Pellegrini}}, \bibinfo {author} {\bibfnamefont {R.}~\bibnamefont
  {C\^ot\'e}}, \bibinfo {author} {\bibfnamefont {M.~D.}\ \bibnamefont {Lukin}},
  \ and\ \bibinfo {author} {\bibfnamefont {S.~F.}\ \bibnamefont {Yelin}},\
  }\href {\doibase 10.1103/PhysRevA.78.021402} {\bibfield  {journal} {\bibinfo
  {journal} {Phys. Rev. A}\ }\textbf {\bibinfo {volume} {78}},\ \bibinfo
  {pages} {021402} (\bibinfo {year} {2008})}\BibitemShut {NoStop}%
\bibitem [{\citenamefont {Zhang}(2024)}]{10.1063/5.0183063}%
  \BibitemOpen
  \bibfield  {author} {\bibinfo {author} {\bibfnamefont {J.}~\bibnamefont
  {Zhang}},\ }\href {\doibase 10.1063/5.0183063} {\bibfield  {journal}
  {\bibinfo  {journal} {J. Chem. Phys.}\ }\textbf {\bibinfo {volume} {160}},\
  \bibinfo {pages} {024104} (\bibinfo {year} {2024})}\BibitemShut {NoStop}%
\bibitem [{\citenamefont {Zhang}\ \emph {et~al.}(2024)\citenamefont {Zhang},
  \citenamefont {Deng}, \citenamefont {Niu},\ and\ \citenamefont
  {Gong}}]{PhysRevA.109.023109}%
  \BibitemOpen
  \bibfield  {author} {\bibinfo {author} {\bibfnamefont {J.}~\bibnamefont
  {Zhang}}, \bibinfo {author} {\bibfnamefont {L.}~\bibnamefont {Deng}},
  \bibinfo {author} {\bibfnamefont {Y.}~\bibnamefont {Niu}}, \ and\ \bibinfo
  {author} {\bibfnamefont {S.}~\bibnamefont {Gong}},\ }\href {\doibase
  10.1103/PhysRevA.109.023109} {\bibfield  {journal} {\bibinfo  {journal}
  {Phys. Rev. A}\ }\textbf {\bibinfo {volume} {109}},\ \bibinfo {pages}
  {023109} (\bibinfo {year} {2024})}\BibitemShut {NoStop}%
\bibitem [{\citenamefont {Rangelov}\ and\ \citenamefont
  {Vitanov}(2012)}]{PhysRevA.85.043407}%
  \BibitemOpen
  \bibfield  {author} {\bibinfo {author} {\bibfnamefont {A.~A.}\ \bibnamefont
  {Rangelov}}\ and\ \bibinfo {author} {\bibfnamefont {N.~V.}\ \bibnamefont
  {Vitanov}},\ }\href {\doibase 10.1103/PhysRevA.85.043407} {\bibfield
  {journal} {\bibinfo  {journal} {Phys. Rev. A}\ }\textbf {\bibinfo {volume}
  {85}},\ \bibinfo {pages} {043407} (\bibinfo {year} {2012})}\BibitemShut
  {NoStop}%
\bibitem [{\citenamefont {Nedaee-Shakarab}\ \emph {et~al.}(2017)\citenamefont
  {Nedaee-Shakarab}, \citenamefont {Saadati-Niari},\ and\ \citenamefont
  {Zolfagharpour}}]{PhysRevC.96.044619}%
  \BibitemOpen
  \bibfield  {author} {\bibinfo {author} {\bibfnamefont {B.}~\bibnamefont
  {Nedaee-Shakarab}}, \bibinfo {author} {\bibfnamefont {M.}~\bibnamefont
  {Saadati-Niari}}, \ and\ \bibinfo {author} {\bibfnamefont {F.}~\bibnamefont
  {Zolfagharpour}},\ }\href {\doibase 10.1103/PhysRevC.96.044619} {\bibfield
  {journal} {\bibinfo  {journal} {Phys. Rev. C}\ }\textbf {\bibinfo {volume}
  {96}},\ \bibinfo {pages} {044619} (\bibinfo {year} {2017})}\BibitemShut
  {NoStop}%
\bibitem [{\citenamefont {Wellnitz}\ \emph {et~al.}(2020)\citenamefont
  {Wellnitz}, \citenamefont {Sch\"utz}, \citenamefont {Whitlock}, \citenamefont
  {Schachenmayer},\ and\ \citenamefont {Pupillo}}]{PhysRevLett.125.193201}%
  \BibitemOpen
  \bibfield  {author} {\bibinfo {author} {\bibfnamefont {D.}~\bibnamefont
  {Wellnitz}}, \bibinfo {author} {\bibfnamefont {S.}~\bibnamefont {Sch\"utz}},
  \bibinfo {author} {\bibfnamefont {S.}~\bibnamefont {Whitlock}}, \bibinfo
  {author} {\bibfnamefont {J.}~\bibnamefont {Schachenmayer}}, \ and\ \bibinfo
  {author} {\bibfnamefont {G.}~\bibnamefont {Pupillo}},\ }\href {\doibase
  10.1103/PhysRevLett.125.193201} {\bibfield  {journal} {\bibinfo  {journal}
  {Phys. Rev. Lett.}\ }\textbf {\bibinfo {volume} {125}},\ \bibinfo {pages}
  {193201} (\bibinfo {year} {2020})}\BibitemShut {NoStop}%
\bibitem [{\citenamefont {Ciamei}\ \emph {et~al.}(2017)\citenamefont {Ciamei},
  \citenamefont {Bayerle}, \citenamefont {Chen}, \citenamefont {Pasquiou},\
  and\ \citenamefont {Schreck}}]{PhysRevA.96.013406}%
  \BibitemOpen
  \bibfield  {author} {\bibinfo {author} {\bibfnamefont {A.}~\bibnamefont
  {Ciamei}}, \bibinfo {author} {\bibfnamefont {A.}~\bibnamefont {Bayerle}},
  \bibinfo {author} {\bibfnamefont {C.-C.}\ \bibnamefont {Chen}}, \bibinfo
  {author} {\bibfnamefont {B.}~\bibnamefont {Pasquiou}}, \ and\ \bibinfo
  {author} {\bibfnamefont {F.}~\bibnamefont {Schreck}},\ }\href {\doibase
  10.1103/PhysRevA.96.013406} {\bibfield  {journal} {\bibinfo  {journal} {Phys.
  Rev. A}\ }\textbf {\bibinfo {volume} {96}},\ \bibinfo {pages} {013406}
  (\bibinfo {year} {2017})}\BibitemShut {NoStop}%
\bibitem [{\citenamefont {Zhang}\ and\ \citenamefont {Dou}(2021)}]{Zhang_2021}%
  \BibitemOpen
  \bibfield  {author} {\bibinfo {author} {\bibfnamefont {J.}~\bibnamefont
  {Zhang}}\ and\ \bibinfo {author} {\bibfnamefont {F.}~\bibnamefont {Dou}},\
  }\href {\doibase 10.1088/1367-2630/abffff} {\bibfield  {journal} {\bibinfo
  {journal} {New J. Phys.}\ }\textbf {\bibinfo {volume} {23}},\ \bibinfo
  {pages} {063001} (\bibinfo {year} {2021})}\BibitemShut {NoStop}%
\bibitem [{\citenamefont {Malinovsky}\ and\ \citenamefont
  {Tannor}(1997)}]{PhysRevA.56.4929}%
  \BibitemOpen
  \bibfield  {author} {\bibinfo {author} {\bibfnamefont {V.~S.}\ \bibnamefont
  {Malinovsky}}\ and\ \bibinfo {author} {\bibfnamefont {D.~J.}\ \bibnamefont
  {Tannor}},\ }\href {\doibase 10.1103/PhysRevA.56.4929} {\bibfield  {journal}
  {\bibinfo  {journal} {Phys. Rev. A}\ }\textbf {\bibinfo {volume} {56}},\
  \bibinfo {pages} {4929} (\bibinfo {year} {1997})}\BibitemShut {NoStop}%
\bibitem [{\citenamefont {Shore}\ \emph {et~al.}(1991)\citenamefont {Shore},
  \citenamefont {Bergmann}, \citenamefont {Oreg},\ and\ \citenamefont
  {Rosenwaks}}]{PhysRevA.44.7442}%
  \BibitemOpen
  \bibfield  {author} {\bibinfo {author} {\bibfnamefont {B.~W.}\ \bibnamefont
  {Shore}}, \bibinfo {author} {\bibfnamefont {K.}~\bibnamefont {Bergmann}},
  \bibinfo {author} {\bibfnamefont {J.}~\bibnamefont {Oreg}}, \ and\ \bibinfo
  {author} {\bibfnamefont {S.}~\bibnamefont {Rosenwaks}},\ }\href {\doibase
  10.1103/PhysRevA.44.7442} {\bibfield  {journal} {\bibinfo  {journal} {Phys.
  Rev. A}\ }\textbf {\bibinfo {volume} {44}},\ \bibinfo {pages} {7442}
  (\bibinfo {year} {1991})}\BibitemShut {NoStop}%
\bibitem [{\citenamefont {B\"uchler}\ \emph {et~al.}(2007)\citenamefont
  {B\"uchler}, \citenamefont {Demler}, \citenamefont {Lukin}, \citenamefont
  {Micheli}, \citenamefont {Prokof'ev}, \citenamefont {Pupillo},\ and\
  \citenamefont {Zoller}}]{PhysRevLett.98.060404}%
  \BibitemOpen
  \bibfield  {author} {\bibinfo {author} {\bibfnamefont {H.~P.}\ \bibnamefont
  {B\"uchler}}, \bibinfo {author} {\bibfnamefont {E.}~\bibnamefont {Demler}},
  \bibinfo {author} {\bibfnamefont {M.}~\bibnamefont {Lukin}}, \bibinfo
  {author} {\bibfnamefont {A.}~\bibnamefont {Micheli}}, \bibinfo {author}
  {\bibfnamefont {N.}~\bibnamefont {Prokof'ev}}, \bibinfo {author}
  {\bibfnamefont {G.}~\bibnamefont {Pupillo}}, \ and\ \bibinfo {author}
  {\bibfnamefont {P.}~\bibnamefont {Zoller}},\ }\href {\doibase
  10.1103/PhysRevLett.98.060404} {\bibfield  {journal} {\bibinfo  {journal}
  {Phys. Rev. Lett.}\ }\textbf {\bibinfo {volume} {98}},\ \bibinfo {pages}
  {060404} (\bibinfo {year} {2007})}\BibitemShut {NoStop}%
\bibitem [{\citenamefont {Baranov}\ \emph {et~al.}(2012)\citenamefont
  {Baranov}, \citenamefont {Dalmonte}, \citenamefont {Pupillo},\ and\
  \citenamefont {Zoller}}]{Baranov2012}%
  \BibitemOpen
  \bibfield  {author} {\bibinfo {author} {\bibfnamefont {M.~A.}\ \bibnamefont
  {Baranov}}, \bibinfo {author} {\bibfnamefont {M.}~\bibnamefont {Dalmonte}},
  \bibinfo {author} {\bibfnamefont {G.}~\bibnamefont {Pupillo}}, \ and\
  \bibinfo {author} {\bibfnamefont {P.}~\bibnamefont {Zoller}},\ }\href
  {\doibase 10.1021/cr2003568} {\bibfield  {journal} {\bibinfo  {journal}
  {Chem. Rev.}\ }\textbf {\bibinfo {volume} {112}},\ \bibinfo {pages} {5012}
  (\bibinfo {year} {2012})}\BibitemShut {NoStop}%
\bibitem [{\citenamefont {Gorshkov}\ \emph {et~al.}(2011)\citenamefont
  {Gorshkov}, \citenamefont {Manmana}, \citenamefont {Chen}, \citenamefont
  {Ye}, \citenamefont {Demler}, \citenamefont {Lukin},\ and\ \citenamefont
  {Rey}}]{PhysRevLett.107.115301}%
  \BibitemOpen
  \bibfield  {author} {\bibinfo {author} {\bibfnamefont {A.~V.}\ \bibnamefont
  {Gorshkov}}, \bibinfo {author} {\bibfnamefont {S.~R.}\ \bibnamefont
  {Manmana}}, \bibinfo {author} {\bibfnamefont {G.}~\bibnamefont {Chen}},
  \bibinfo {author} {\bibfnamefont {J.}~\bibnamefont {Ye}}, \bibinfo {author}
  {\bibfnamefont {E.}~\bibnamefont {Demler}}, \bibinfo {author} {\bibfnamefont
  {M.~D.}\ \bibnamefont {Lukin}}, \ and\ \bibinfo {author} {\bibfnamefont
  {A.~M.}\ \bibnamefont {Rey}},\ }\href {\doibase
  10.1103/PhysRevLett.107.115301} {\bibfield  {journal} {\bibinfo  {journal}
  {Phys. Rev. Lett.}\ }\textbf {\bibinfo {volume} {107}},\ \bibinfo {pages}
  {115301} (\bibinfo {year} {2011})}\BibitemShut {NoStop}%
\end{thebibliography}%

\end{document}